\documentclass[%
 reprint,
 amsmath,amssymb,
 aps,
]{revtex4-2}

\usepackage{graphicx}% Include figure files
\usepackage{dcolumn}% Align table columns on decimal point
\usepackage{bm}% bold math
\usepackage{orcidlink}%

\begin{document}

%\preprint{}

\title{Remarks on ``The Modification of Feynman Diagrams in Curved Space-Time"}


\author{Noah M. MacKay\,\orcidlink{0000-0001-6625-2321}}
\email{noah.mackay@uni-potsdam.de}
%\thanks{}
\affiliation{Institut für Physik und Astronomie, Universität Potsdam\\
        Karl-Liebknecht-Straße 24/25, 14476 Potsdam, Germany}


\date{\today}

\begin{abstract}
The arXiv preprint 2411.15164v1 \cite{Li:2024ltx} discusses how a weak Schwarzschild spacetime modifies Feynman diagram calculations and the governing equations of scalar and spinor particles. This short review offers an extended analysis on concerning points involving massless particles and limitations of weak-field approximation. Written for the Universit\"at Potsdam lecture \textit{New Developments in Astrophysics, Winter Semester 2024/25}.
\end{abstract}

%\keywords{}

\maketitle

%\tableofcontents

\section{Introduction}

Ref. \cite{Li:2024ltx} is an insightful contribution to science in the context of better understanding particle interactions in a Newtonian gravitational well. Weak field approximation and Local Minkowski Coordinates are valid for this scenario, which may also extend to particle interactions in a globally curved but locally flat fluid. However, cases where a strong Schwarzschild spacetime is essential are overlooked for the sake of simplification. Ultra-relativistic characteristics also cause a concerning point worth discussing. 

\section{$B(k)$ for Massless Particles}

For a Schwarzschild metric, the geometric probability modifier $B(k)$ ($k$ is the 3-momentum) is defined in full as
\begin{equation}\label{Bk}
    B(k)=\sqrt{\frac{k(r_0)}{k(r)}}=\sqrt[4]{\frac{E^2-m^2(1-r_S/r_0)}{E^2-m^2(1-r_s/r)}}.
\end{equation}
where both numerator and denominator show a modified dispersion relation in terms of radial displacement from the Schwarzschild radius $r_s\equiv 2GM$ ($c=1)$. The numerator is fixed at an observer displacement $r_0>r_s$, and the denominator is variable with $r>r_s$, imposing the weak field approximation ($r_0,r\gg r_s$ yields the original dispersion relation). However, massless particles (e.g., photons and gluons) have unitary probability given the formula:
\begin{equation}
   \lim_{m\rightarrow0} B(k)=\sqrt[4]{\frac{E^2}{E^2}}=1,
\end{equation}
which is the same scenario for observing any-mass particles at $r=r_0$.  This suggests that observing distant massless particles on any spacetime is the same as observing nearby massless particles on a local Minkowski spacetime, i.e. as though the massless particle's probability were unmodified by gravity. This is at odds with the known effects a curved manifold has on null geodesics, where the massless particle probability would intuitively be influenced by the global curvature.

A possible modification may be revising the massless dispersion relation in the form of $k^2=E^2-Ak^\alpha f(r)$, where $f(r)=(1-r_s/r)$, $|A|$ is magnitude of dispersion, and $\alpha\geq0$. Lorentz invariance is not violated under $|A|\ll1$ and $\alpha=0$ \cite{LIGOScientific:2017bnn}, which assigns \textit{effective} mass $\mu$ to a massless field. While this may cause fundamental ramifications, such as deviations from Maxwell equations for "massive photons," these deviations may be neglected if $\mu$ is small. Effective masses are also important in screening away singularities in Feynman exchange channels, especially in kinematic regimes where interacting particles act ultra-relativistically \cite{Zhang:1997ej, Arnold:2003zc}.

\section{Place-marking the Weak-to-Strong Transition Point}

The results in the main text \cite{Li:2024ltx} are valid, given that our manifold has a weak Schwarzschild geometry. While this is applicable for Newtonian objects with surface radii $R>r_s$, it is ambiguous if the Schwarzschild black hole is the host gravitating object. This introduces the question, generally, where can a weak-to-strong transition point be marked given $f(r)=1-r_s/r$?

\begin{figure}[h!]
    \centering
    \includegraphics[width=80mm]{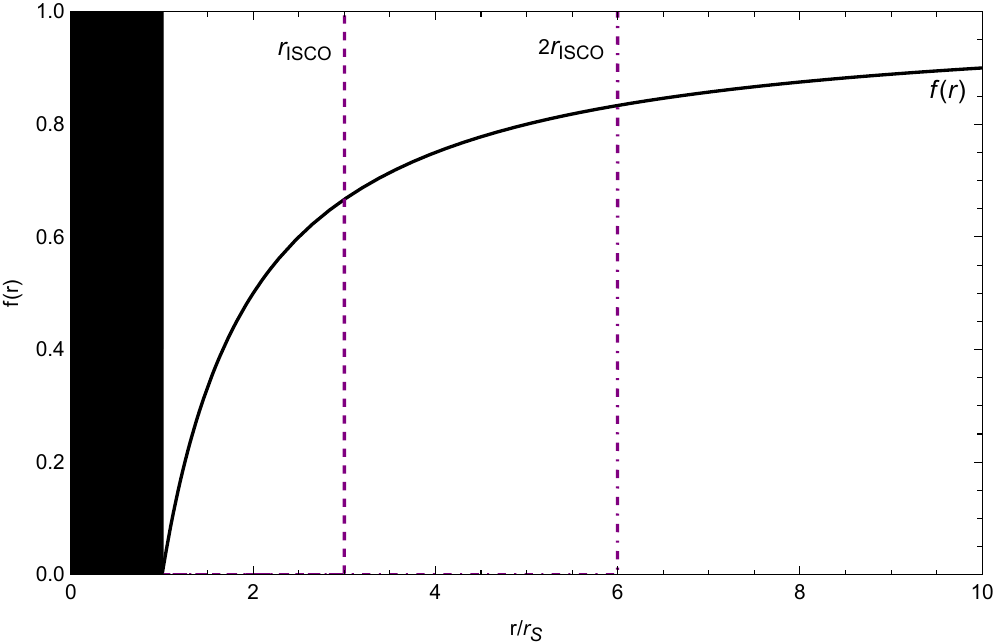}
    \caption{ \label{fig:plot} Map of a potential landscape drawn by $f(r)$ in thin black. The black boundary between $r\in[0,1]r_s$ is the Schwarzschild black hole. Purple lines respectively mark $r_\mathrm{ISCO}$ and $2r_{\mathrm{ISCO}}$. }
\end{figure}

Naively, with the innermost stable circular orbit on a Schwarzschild geometry marked at $r_{\mathrm{ISCO}}=3r_s$, the "weak-to-strong marker" at $r_1=2r_{\mathrm{ISCO}}$ would be a safe bet. I.e., radial displacements of $r>2r_{\mathrm{ISCO}}$ from the central object allow for the weak Schwarzschild approximation. This is illustrated in Figure \ref{fig:plot} with a potential landscape drawn by $f(r)=(1-r_s/r)$ versus $r/r_s$. A critical location along $r$ where a hypothetical marble begins a non-linear, downhill roll towards $r_s$ should be where $r_1$ is marked.

\section{Inclusion of Vector and Tensor Particles}

The preprint focuses on scalar (spin-0) particles and spinor (spin-1/2) particles, whose governing equations are generalized for any geometry and mass is kept as non-zero. Expanding its application to other particle types (i.e., vector and tensor) are explored here.

Vector and tensor fields in QFT respectively correspond to massless spin-1 particles (e.g., photons and gluons) and massless spin-2 particles (e.g., gravitons). This talking point is in continuation from the first point ($B(k)$ for $m\rightarrow0$), however unitary probability is treated as matter of fact.

The source of vector particles under the Minkowski metric are flux tensors (i.e., Faraday flux and gluon field flux), which generate a 4-potential that propagates as the particle under the Lorentz and Coulomb gauges \cite{griffiths}:
\begin{eqnarray}
&&\mathrm{QED}:~~F^{\mu\nu}=\partial^\mu A^\nu-\partial^\nu A^\mu\nonumber\\
 &&~~~~~~~~~~~~\implies \Box A^\mu=0, \nonumber\\
        &&\mathrm{QCD}:~~\mathcal{G}^{\mu\nu}_\mathrm{a}=\partial^\mu\mathcal{A}_\mathrm{a}^\nu-\partial^\nu\mathcal{A}_\mathrm{a}^\mu+g_sf^{\mathrm{abc}}\mathcal{A}^\mu_\mathrm{b}\mathcal{A}^\nu_{\mathrm{c} }\nonumber\\
        &&~~~~~~~~~~~~\implies \Box\mathcal{A}^\mu_a=0.
\end{eqnarray}
The QED 4-potential propagates a photon; the QCD 4-potential propagates a gluon, where the extra term $g_sf^{\mathrm{abc}}\mathcal{A}^\mu_\mathrm{b}\mathcal{A}^\nu_{\mathrm{c} }$ enforces self-interaction among gluons. Also for gluons, the index "a" is a color-index, which is separate from spacetime indices.

For both particles, $\Box=\eta_{\alpha\beta}\partial^\alpha\partial^\beta$, and their vector wave-functions are proportional to a spin-polarization vector: 
\begin{equation}
  ( A_\mu,~\mathcal{A}_\mu^a)\propto\varepsilon_\mu.
\end{equation}
Their solutions depend on 2 degrees of freedom imposed by the Lorentz and Coulomb gauges (i.e., the two polarization states orthogonal to propagation), which is reflected in the spin vector $\varepsilon_\mu$.

There is, as of now, no widely accepted QFT involving tensor particles. For GR, the particle mediator is the graviton, which ideally is in wave-particle duality with GWs. For tensor fields in a weak field approximation, where $g_{\mu\nu}=\eta_{\mu\nu}+h_{\mu\nu}$, a wave equation for the perturbation $h_{\mu\nu}$ ultimately comes from very small changes in the Ricci tensor:
\begin{eqnarray}
&&\delta R_{\mu\nu}=-\frac{1}{2}\Box h_{\mu\nu}\nonumber\\
&&\rightarrow\quad\Box h_{\mu\nu}=0~~\mathrm{with}~~\delta R_{\mu\nu}=0.
\end{eqnarray}

As like for vector particles, $\Box=\eta_{\alpha\beta}\partial^\alpha\partial^\beta$ applies here. A solution to the reduced equation depends, therefore, on a spin polarization tensor \cite{Feynman:2002}:
\begin{equation}
   h_{\mu\nu}\propto\varepsilon_{\mu\nu},
\end{equation}
whose degrees of freedom vastly reduce from 10 to 2 via the gauge $\partial_\beta h^\beta _\alpha-(\partial_\alpha h^\beta_\beta)/2=0$. The 2 physical degrees of freedom are collectively called the traceless-transverse (TT) gauge, which is akin to the photon's and gluon's Lorentz and Coulomb gauges. Should gravitons propagate as GWs, they have 2 degrees of freedom despite being a spin-2 particle. This makes tensor particles quite relatable to the vector particles, provided they are massless. Otherwise, any non-zero mass assigned to the graviton increases the degrees of freedom from 2 to 5 via $2s+1$ in 3+1 spacetime. The associating "massive gravity" theory applies to minuscule dispersion in GWs upon contact with detectors, determining a threshold graviton mass of $m_g=7.7\times10^{-23}~\mathrm{eV}$ \cite{LIGOScientific:2017bnn}.

As the vector and tensor wave-functions are governed by the d'Alembert operator $\Box$, the ansatz of using the Laplace-Beltrami generalization is utilized:
\begin{eqnarray}
        &&\mathrm{photons}:~ \frac{1}{\sqrt{-g}}\partial_\alpha(\sqrt{-g}g^{\alpha\beta}\partial_\beta)A_\mu=0, \nonumber\\
        &&\mathrm{gluons}:~ \frac{1}{\sqrt{-g}}\partial_\alpha(\sqrt{-g}g^{\alpha\beta}\partial_\beta)\mathcal{A}_\mu^a=0,\nonumber\\
        &&\mathrm{GWs/gravitons:}~\frac{1}{\sqrt{-g}}\partial_\alpha(\sqrt{-g}g^{\alpha\beta}\partial_\beta)h_{\mu\nu}=0.
\end{eqnarray}
Now the question of the correction's validity remains.

For photons, gamma ray interactions with interstellar media that may have global curvatures would encourage geometric corrections to the photon equation. Gravitational lensing and redshift in a weak Schwarschild geometry may also be explained through this generalization.

Gluons do not exist in the vacuum as free particles, for they contain color charge. The scenario where both quarks and gluons are quasi-free involves a deconfined phase that acts as a near-perfect fluid with a short lifespan \cite{MacKay:2022uxo}. In this phase (the quark-gluon plasma), curvature corrections to both gluon and Dirac equations may be applicable, provided the quark-gluon plasma can curve spacetime -- if created in a Minkowski vacuum.

For GR-mediating gravitons, their elusiveness scrutinizes their relevance in particle physics beyond string theory. In the context of GWs, propagation into a weak Schwarzschild geometry involves contact with Newtonian objects (e.g. the Earth via LVK observatories). Scattering amplitudes between TT-gauged GWs and scalar/spinor particles can be calculated (see Ref. \cite{MacKay:2024sgw} for LIGO-like Compton scattering), which may be insightful for gravitational effects in the quantum scale.

%\pagebreak

\end{document}